\documentclass{article}
\usepackage{spconf,amsmath,graphicx}
\usepackage{enumitem}
\usepackage{adjustbox}
\usepackage[page]{appendix}
\PassOptionsToPackage{hyphens}{url}\usepackage{hyperref}
\usepackage{listings}
\hypersetup{colorlinks=true}

\usepackage{amsfonts}
\usepackage{booktabs,multirow}
\usepackage{adjustbox}
\usepackage{algorithm}
\usepackage{algpseudocode}
\usepackage{color}
\usepackage{cite}
\usepackage{xcolor}

\usepackage{caption}
\newcommand{\mat}[1]{\boldsymbol{\mathbf{#1}}} 

\newcommand{\footnoteurl}[1]{\footnote{\scriptsize \url{#1}}}

\title{Augmenting Transformer-Transducer Based Speaker Change Detection With Token-Level Training Loss}

\name{Guanlong Zhao, Quan Wang, Han Lu, Yiling Huang, Ignacio Lopez Moreno}
\address{Google LLC, USA \\
\vspace{-4mm}
\\\texttt{\normalsize \{\href{mailto:guanlongzhao@google.com}{guanlongzhao},\href{mailto:quanw@google.com}{quanw},\href{mailto:luha@google.com}{luha},\href{mailto:yilinghuang@google.com}{yilinghuang},\href{mailto:elnota@google.com}{elnota}\}@google.com}}

\begin{document}
\ninept
\maketitle
\begin{abstract}
\vspace{6pt}
In this work we propose a novel token-based training strategy that improves Transformer-Transducer (T-T) based speaker change detection (SCD) performance. The conventional T-T based SCD model loss optimizes all output tokens equally. Due to the sparsity of the speaker changes in the training data, the conventional T-T based SCD model loss leads to sub-optimal detection accuracy. To mitigate this issue, we use a customized edit-distance algorithm to estimate the token-level SCD false accept (FA) and false reject (FR) rates during training and optimize model parameters to minimize a weighted combination of the FA and FR, focusing the model on accurately predicting speaker changes. We also propose a set of evaluation metrics that align better with commercial use cases. Experiments on a group of challenging real-world datasets show that the proposed training method can significantly improve the overall performance of the SCD model with the same number of parameters.
\end{abstract}

\begin{keywords}
Speaker change detection, speaker turn detection, Transformer-Transducer, minimum Bayes risk training
\end{keywords}

\vspace{-6pt}
\section{Introduction}
\label{sec:intro}
\vspace{-2pt}
Speaker change detection (SCD) or speaker turn detection  \cite{chen1998speaker,ajmera2004robust,yin2017speaker,xia2022turn} is the process of identifying the speaker change points in a multi-speaker continuous conversational input audio stream. SCD has broad applications in enhancing speaker diarization accuracy \cite{xia2022turn,wang2022highly}, improving automatic speech recognition quality \cite{sari2020auxiliary}, and generating line breaks in captions to augment readability and accessibility \cite{donabauer2021making}.

Conventionally, SCD is achieved by using a neural network to map acoustic features (e.g., spectral/cepstral coefficients of various flavors \cite{hruz2017convolutional,yin2017speaker,yin2018neural}) or biometric features \cite{aronowitz2020context} (e.g., speaker embeddings \cite{dehak2010front,snyder2018x,variani2014deep,ge2e}) to a frame or segment level binary prediction --- yes/no speaker change. The neural network is generally trained by minimizing the binary cross entropy loss between the ground-truth speaker change labels and the predictions. These conventional approaches have various limitations. First, they require accurate timing information of the speaker change point, which is difficult since marking speaker change points is a highly subjective process for human annotators. Second, the methods that use purely acoustic information ignore rich semantic information in the audio signal. For example, by only looking at the text transcript of the conversation ``How are you I'm good'', we can confidently conjecture there is a speaker change between ``How are you'' and ``I'm good''. Third, the methods that use speaker embeddings for SCD utilize sensitive biometric information that can be exploited for unintended purposes. Thus, they are sub-optimal from a privacy point of view~\cite{de2017europe}.

To mitigate the aforementioned issues, previously we have proposed an SCD model using a Transformer-Transducer (T-T) \cite{xia2022turn}. Specifically, we augment the text transcription of the spoken utterance with a special speaker turn token \texttt{<st>}, and then train the model to output both regular text tokens and the special speaker turn token. This model does not need accurate timestamps for training since the T-T model is trained in a seq2seq fashion and does not need forced-alignment to provide training targets. The model also utilizes both acoustic and linguistic information in the input audio.

Speaker turns are relatively sparse compared to regular spoken words. Based on estimates on the training data we use, regular spoken words appear 40+ times more frequently than speaker turns. The conventional T-T training loss maximizes the log probability of the entire output sequence, including both spoken words and the special speaker turn token. The sparsity of speaker changes in the training data leads to them being de-emphasized, resulting in high error rates in the inference. In the ASR research community, minimum Bayes risk (MBR) training \cite{vesely2013sequence} is a widely used technique to improve model performance. One common variation of this technique is the Edit-distance based MBR (EMBR)\footnote{\scriptsize Also known as Minimum Word Error Rate (MWER).} technique, which optimizes model parameters to minimize the expected word error rate \cite{shannon2017optimizing}. Inspired by the EMBR technique, we propose to supplement the original T-T training loss with additional token-based penalty terms to minimize the expected recognition error of the speaker turns. During training, we apply a constrained edit distance algorithm to identify the speaker change false acceptance (FA) and false rejection (FR) errors in the N-best hypotheses, and then penalize the training loss to steer the gradient away from the high FA/FR region.

The contributions of this work are two-fold. First, we propose a training loss for SCD that directly minimizes the expected token-level FA and FR rates, and results in improved SCD performance. Second, we define a new set of SCD evaluation metrics and demonstrate that they can better reflect the model quality than previous metrics \cite{bredin2017pyannote} on a group of diverse test sets.

\vspace{-6pt}
\section{System Description}
\label{sec:sys}

\vspace{-2pt}
\subsection{Baseline SCD model}
\label{sec:sys:baseline}

The recurrent neural network transducer (RNN-T)~\cite{graves2012sequence} is an ASR model architecture that can be trained end-to-end with the RNN-T loss. An RNN-T model includes an audio encoder, a label encoder, and a joint network that produces the output softmax distribution over a predefined vocabulary. We adopt the Transformer Transducer ~\cite{zhang2020transformer}, a variant of the RNN-T architecture, as the speaker change detection model for its advantages of faster inference speed and better handling of long-form deletion issues. We use LSTM layers as the label encoder, and fully connected layers as the joint network.

To create training targets, we add a special speaker turn token \texttt{<st>} between two different speakers' transcripts (e.g. ``hello how are you \texttt{<st>} I am good \texttt{<st>}'') to model speaker turns during training. This is inspired by \cite{shafey2019joint} that adds speaker roles as part of the transcript (e.g. ``hello how are you \texttt{<spk:dr>} I am good \texttt{<spk:pt>}''). Compared with audio-only SCD models~\cite{yin2017speaker,yin2018neural}, this model can potentially utilize the language semantics as a signal for speaker segmentation. T-T is trained in a seq2seq fashion, where the input sequence contains log-Mel filterbank energy features, and the output sequence contains the transcript that includes both transcript texts and the special speaker turn tokens. For inference, we perform a beam-search with the T-T SCD model's softmax outputs, and identify the speaker turn tokens. We use the timestamps of the predicted speaker turn tokens in the evaluation.

\vspace{-6pt}
\subsection{Token-based training loss}
\label{sec:sys:proposed}


To focus modeling capacity on the SCD task, we augment the training loss with an additional token-based SCD penalty. On a high level, we first construct a T-T model that takes audio as input, and outputs the speaker turn augmented transcriptions. We optimize the T-T model by maximizing the log probability and following the process described in Sec. \ref{sec:sys:baseline}. We then warm-start a new T-T model with the model trained in the previous step and fine-tune the new model on the same training data with the following steps: (1) for each training utterance, perform a beam search to get the N-best hypotheses associated with their corresponding probability scores, which is how likely the hypotheses appear based on the existing model parameters; (2) compute the token-level FA and FR from the N-best hypotheses; (3) fine-tune the model with a loss function that is a weighted sum of the log probability and token-level FA and FR rates.

\vspace{-6pt}
\subsubsection{Token-level FA and FR}

Mathematically, let $M$ be the number of training samples and $N$ be the number of hypotheses per training sample; let $\mat{H}_{ij}$ be the $j$-th hypothesis of the $i$-th training sample, where $i \in [1, M]$ and $j \in [1, N]$; let $\mat{P}_{ij}$ be the probability score associated with $\mat{H}_{ij}$ given by the model; let $\mat{R}_{ij}$ be the corresponding reference transcription. We first compute a customized minimum edit distance alignment \cite{jurafsky2008speech} between all $\mat{H}_{ij}$ and $\mat{R}_{ij}$. The idea of the customized minimum edit distance alignment is to only allow substitutions among regular spoken words, and each special speaker turn token prediction \texttt{<st>} can only be one of \{\texttt{correct, deleted, inserted}\}. A substitution error between a regular spoken word token and the special speaker turn token is ill-defined, thus, not allowed in the customized minimum edit distance alignment. To achieve this, the edit distance algorithm applies the following costs for its optimization,

\begin{equation}
\label{eq:sub_costs}
\text{sub-cost}(r, h)=
\begin{cases}
{0, }& {\text{If } r=h;} \\
{1, }& \text{If }r\ne h\ne\text{\texttt{<st>}}; \\
{+\infty, }& \text{Otherwise}.
\end{cases}
\end{equation}

\begin{equation}
\label{eq:ins_del_costs}
\text{ins/del-cost}(token)=
\begin{cases}
{k, }& \text{If }token=\text{\texttt{<st>}}; \\
{1, }& \text{Otherwise}.
\end{cases}
\end{equation}

Here, $r$ and $h$ are tokens in $\mat{R}_{ij}$ and $\mat{H}_{ij}$, respectivly. The constant $k\geq1$ controls the tolerance of the offset in predicting \texttt{<st>}. If $k=1$, we expect an exact match between the reference and predicted \texttt{<st>} tokens. If $k>1$, we allow a maximum offset of $\lfloor k\rfloor$ tokens between a pair of reference and predicted \texttt{<st>} tokens for them to be considered as correctly aligned, offering some tolerance on annotation errors.

\vspace{-6pt}
\subsubsection{Training loss}

Based on the optimal alignment obtained from the customized minimum edit distance, we identify the number of speaker turn token insertions (denoted as $\mat{FA}_{ij}$) and deletions ($\mat{FR}_{ij}$) as well as the number of spoken word errors $\mat{W}_{ij}$ in $\mat{H}_{ij}$. We compute the per sample token-level loss as

\begin{equation}
\label{eq:per_hyp_loss}
\mat{L}_{ij} = \mat{P}_{ij} \cdot \frac{\alpha \mat{W}_{ij} + \beta \mat{FA}_{ij} + \gamma \mat{FR}_{ij}}{\mat{Q}_{ij}},
\end{equation}

\noindent where $\{\alpha, \beta, \gamma\}$ control the relative strength of each subcomponent, and $\mat{Q}_{ij}$ is the total number of tokens in $\mat{R}_{ij}$. We generally set $\beta$ and $\gamma$ to be much larger than $\alpha$ to force the model to reduce the speaker change insertion and deletion rates. In essence, Eq.~(\ref{eq:per_hyp_loss}) is a weighted sum of the expected value of WER, SCD FA, and SCD FR. We compute the final per batch training loss as

\begin{equation}
\label{eq:batch_loss}
L_\text{SCD} = \sum_{i=1}^{M}{\sum_{j=1}^{N}{\mat{L}_{ij}}} - \lambda \log P(\mat{Y}|\mat{X}),
\end{equation}

\noindent where $-\log P(\mat{Y}|\mat{X})$ is the negative log probability of the ground-truth transcription $\mat{Y}$ conditioned on the input acoustic features $\mat{X}$. The regularization term $\lambda$ controls the strength of the negative log probability loss.

\vspace{-6pt}
\section{Evaluation metrics}
\label{sec:metrics}

Conventional evaluation metrics for the speaker change detection task are timestamp-based precision and recall rates \cite{hruz2017convolutional}. A predicted speaker change point is considered as correct if it falls within a temporal window (the ``collar") surrounding a reference speaker change point. Therefore, these metrics are very sensitive to the temporal precision of human annotations, and the precision and recall rates quickly reach zero when the collar approaches zero.

\vspace{-6pt}
\subsection{Interval-based precision and recall}
\label{sec:precision_recall}

\begin{figure}[t!]
    \begin{center}
    	\includegraphics[trim={2.3in 1.65in 2.1in 1.7in},clip,width=3.49in]{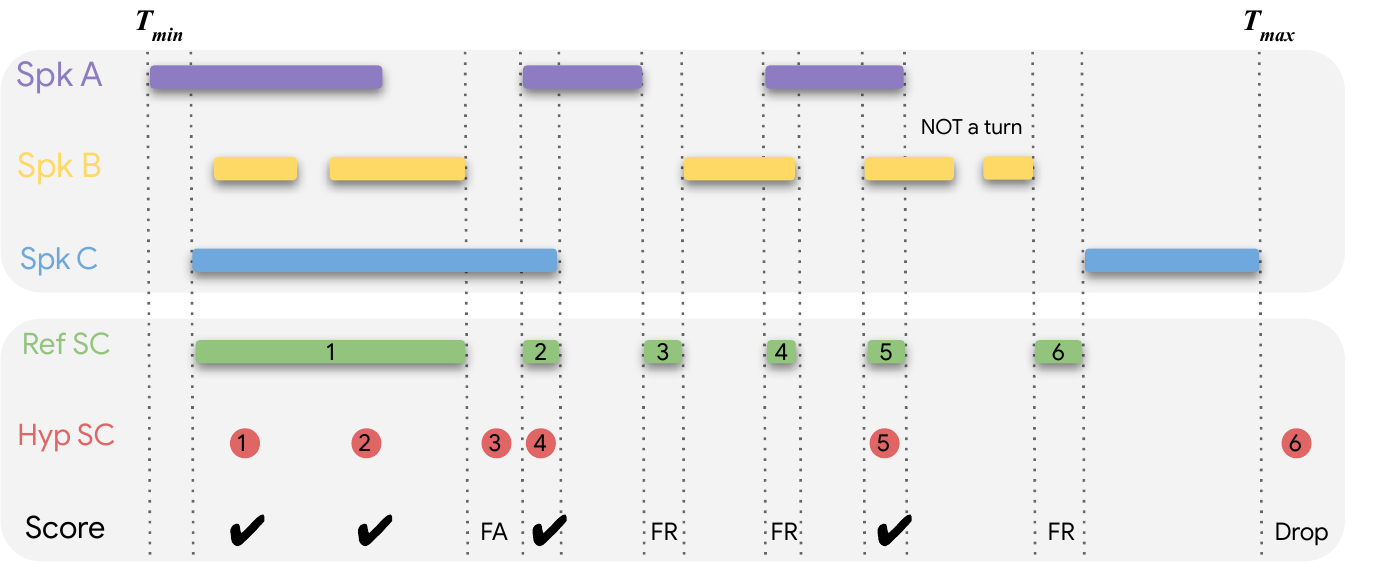}
    	\caption{Illustration of the various components for computing the precision and recall rates. ``Spk A-C" stands for speaker annotations on a conversational utterance. ``Ref SC" is the speaker change intervals (i.e., $\overline{U}$). ``Hyp SC" is the predicted speaker change. ``Score" shows the scoring decision of each prediction and reference.}
    	\label{fig:metrics}
	\end{center}
\end{figure}

In applications such as line breaking in captioning, we argue that speaker changes should be treated as time intervals instead of time-stamped change ``points". For example, in a conversational recording, if speaker A spoke from 0.1-10.5s and speaker B spoke from 10.8-15.3s, the time interval of 10.5-10.8s is where the speaker change happened. If an SCD system predicts a speaker change within this speaker change interval, we should treat this prediction as correct. Following this argument, we propose a new way to compute SCD precision and recall rates by matching the predicted speaker changes to the ground-truth speaker change intervals. We assume the following when computing these metrics: (1) we use speaker annotations to infer speaker change intervals; (2) the test data have ``dense" speaker annotations. ``Dense" annotation means that the test utterance has speaker labels throughout and there are no large unannotated chunks of audio. As an example, it is considered bad data if the annotator labeled one minute of speech, then skipped 10 minutes, and annotated another two minutes of speech. The metrics are computed as follows (refer to Fig. \ref{fig:metrics} for the visualization of different components).

First, based on the reference speaker annotations, we derive the set of mono-speaker time ranges (denoted as $U$) as well as the min-start ($T_{\text{min}}$) and max-end ($T_{\text{max}}$) time of all speaker annotations. For a densely annotated recording, $T_{\text{min}}$ and $T_{\text{max}}$ should generally be the start and end times of the entire utterance. However, there may be cases where the annotations do not cover the full utterance, for example, to skip initial and trailing non-speech audio.

Within the $[T_{\text{min}}, T_{\text{max}}]$ time range, we take the complement set ($\overline{U}$) of the mono-speaker time ranges ($U$) and treat them as speaker change intervals. $\overline{U}$ includes multi-speaker segments and speaker switching segments. We drop the speaker change predictions that are outside of $[T_{\text{min}}, T_{\text{max}}]$ and do not count them during scoring because it is impossible to know whether the predictions outside of the min-max speaker annotation range are valid or not.

For each speaker change prediction, we match it with the reference intervals in $\overline{U}$. If any of the reference intervals overlaps with the prediction, we mark the prediction as correct, otherwise the prediction is an FA. We compute the \emph{precision} rate as the ratio between the number of correct predictions and the total number of predictions. For each reference speaker change interval, if any of the predictions can be matched with it (i.e., overlap in time, aka ``hit"), we consider that reference speaker change interval as being correctly predicted, otherwise the reference speaker change interval is counted as an FR. We compute the \emph{recall} rate as the ratio between the number of ``hit" speaker change intervals and the total number of speaker change intervals. We note that one can also calculate the \emph{recall} based on the duration of the intervals instead of counting the numbers, which would favor longer speaker change intervals.

When performing the matching between speaker change predictions and the reference intervals, we allow a collar of e.g., 250ms, which is common in computing speaker diarization metrics \cite{fiscus2007rich}.

\vspace{-6pt}
\subsection{Purity and coverage}

Purity and coverage scores \cite{yin2017speaker} are commonly used evaluation metrics in SCD tasks. For each reference speaker segment, we find its most overlapping hypothesis speaker segment (defined by the speaker change predictions) and obtain their intersection. The coverage is computed as the ratio between the duration of the intersection and the reference segment. Purity is the dual metric of coverage by swapping the roles of the reference and hypothesis segments in the calculation. By definition, coverage and purity scores focus on the mono-speaker segments in the audio, and therefore can be viewed as a set of complementary metrics to the precision and recall metrics defined above, which focus on evaluating the SCD performance in the speaker change intervals (i.e., non mono-speaker segments).

\vspace{-6pt}
\section{Experimental setup}
\label{sec:exp}

\vspace{-6pt}
\subsection{Datasets}
\label{sec:exp:data}

We train the T-T model using the Fisher corpus \cite{cieri2004fisher}, the training subset of Callhome American English \cite{canavan1997callhome}, the training subset (Full-corpus-ASR partition) of the AMI corpus \cite{carletta2005ami}, the ICSI corpus \cite{janin2003icsi} training subset (we used the data splits from Kaldi \cite{kaldiicsi}), an internal long-form dataset containing around 4,545 hours of audio, and an internal vendor-provided dataset that contains 80 hours of simulated business meeting recordings constructed by having participants discuss randomly assigned business topics with action items. Training utterances are segmented into 15-second segments with speaker turn tokens added so they can fit into the memory of TPUs \cite{jouppi2017datacenter}.

Details of our evaluation datasets are listed in Table~\ref{table:eval_data}. For the first DIHARD challenge evaluation subset (DIHARD1) \cite{ryant2018first}, we removed all YouTube-derived utterances. For Fisher, we withheld a subset\footnote{\scriptsize \url{https://github.com/google/speaker-id/blob/master/publications/ScdLoss/eval/fisher.txt}} of 172 utterances for testing (not used in training). ``Outbound" and ``Inbound" are vendor-provided call center telephone conversations between call center attendants and customers, initiated by the call center and by customers, respectively. ``Outbound" and ``Inbound" were previously used in \cite{xia2022turn,wang2022highly}.

All internal datasets were collected according to Google's Privacy Principles \cite{privacyprinciples} and abide by Google AI Principles \cite{aiprinciples}.

\begin{table}
\centering
\caption{The test sets. For each set, we show the average number of speaker turns per minute and the average per recording duration.}
\resizebox{3.35in}{!}{%
\begin{tabular}{ccccc}
\toprule
\multirow{2}{*}{Testset} & \multirow{2}{*}{Domain} & \multirow{2}{*}{Dur. (h)} & \multicolumn{2}{c}{Average} \\ \cmidrule(lr){4-5}
                         &                         &                           & Turns/min     & Dur./Rec. (min)\\
                         \midrule
AMI                      & Meeting                 & 9.1                       & 10            & 34          \\
Callhome                 & Telephone               & 1.7                       & 19            & 5           \\
DIHARD1                  & Mixed                   & 16.2                      & 12            & 9           \\
Fisher                   & Telephone               & 28.7                      & 13            & 10          \\
ICSI                     & Meeting                 & 2.8                       & 13            & 55          \\
Inbound                  & Telephone               & 21.0                      & 9             & 5           \\
Outbound                 & Telephone               & 45.6                      & 13            & 6           \\
\bottomrule
\end{tabular}%
}
\label{table:eval_data}
\end{table}

\vspace{-6pt}
\subsection{System configurations}
\label{sec:exp:sys}


\begin{table*}[ht!]
\centering
\caption{Long-form evaluation results. The last column shows the evaluation metrics computed by pooling all test sets together.}
\label{tab:long_results}
\resizebox{6.5in}{!}{%
\begin{tabular}{cccccccccc}
\toprule
Evaluation Metric & System      & AMI & CallHome & DIHARD1 & Fisher & ICSI & Inbound & Outbound & Pooled data \\
                                \cmidrule(lr){1-1} \cmidrule(lr){2-2} \cmidrule(lr){3-9} \cmidrule(lr){10-10}
\multirow{3}{*}{Precision (\%)} & Baseline & 80.9 & 81.0 & 78.7 & 81.8 & 78.7 & 73.0 & 76.3 & 78.1\\
                                & EMBR & \textbf{81.3} & \textbf{82.0} & \textbf{79.8} & \textbf{83.5} & \textbf{79.3} & \textbf{74.3} & \textbf{77.0} & \textbf{79.1}\\
                                & SCD loss   & 79.4 & \textbf{82.0} & 78.8 & 82.6 & 77.8 & 72.8 & 75.1 & 77.6\\
                                \cmidrule(lr){1-1} \cmidrule(lr){2-2} \cmidrule(lr){3-9} \cmidrule(lr){10-10}
\multirow{3}{*}{Recall (\%)}    & Baseline & 64.0 & 50.6 & 49.2 & 62.4 & 54.3 & 62.2 & 50.9 & 55.8\\
                                & EMBR & 64.2 & 53.4 & 49.5 & 71.1 & 53.6 & 71.8 & 53.6 & 60.3\\
                                & SCD loss   & \textbf{68.1} & \textbf{59.1} & \textbf{52.4} & \textbf{75.7} & \textbf{58.7} & \textbf{79.2} & \textbf{58.7} & \textbf{65.2}\\
                                \cmidrule(lr){1-1} \cmidrule(lr){2-2} \cmidrule(lr){3-9} \cmidrule(lr){10-10}
\multirow{3}{*}{\shortstack[c]{F1 (\%)\\ (Precision \& Recall)}}   & Baseline & 71.5 & 62.3 & 60.6 & 70.8 & 64.2 & 67.2 & 61.1 & 65.1\\
                                & EMBR & 71.7 & 64.7 & 61.1 & 76.8 & 64.0 & 73.0 & 63.2 & 68.5\\
                                & SCD loss   & \textbf{73.3} & \textbf{68.7} & \textbf{62.9} & \textbf{79.0} & \textbf{66.9} & \textbf{75.9} & \textbf{65.9} & \textbf{70.9}\\
                                \cmidrule(lr){1-1} \cmidrule(lr){2-2} \cmidrule(lr){3-9} \cmidrule(lr){10-10}
\multirow{3}{*}{Purity (\%)}    & Baseline & 87.4 & 84.3 & 90.3 & 80.5 & 76.9 & 95.0 & 76.7 & 82.7\\
                                & EMBR & 87.6 & 84.1 & 90.5 & 82.7 & 77.0 & 95.3 & 77.1 & 83.5\\
                                & SCD loss   & \textbf{88.5} & \textbf{84.9} & \textbf{91.0} & \textbf{83.5} & \textbf{77.7} & \textbf{95.5} & \textbf{78.3} & \textbf{84.3}\\
                                \cmidrule(lr){1-1} \cmidrule(lr){2-2} \cmidrule(lr){3-9} \cmidrule(lr){10-10}
\multirow{3}{*}{Coverage (\%)}  & Baseline & \textbf{70.0} & \textbf{85.6} & 64.9 & \textbf{80.8} & 79.3 & \textbf{77.1} & 83.4 & \textbf{78.5}\\
                                & EMBR & 70.0 & 85.3 & \textbf{65.1} & 80.6 & \textbf{79.8} & 76.7 & \textbf{83.7} & \textbf{78.5}\\
                                & SCD loss   & 68.7 & 84.7 & 64.7 & 80.2 & 78.9 & 75.0 & 82.4 & 77.5\\
                                \cmidrule(lr){1-1} \cmidrule(lr){2-2} \cmidrule(lr){3-9} \cmidrule(lr){10-10}
\multirow{3}{*}{\shortstack[c]{F1 (\%)\\ (Purity \& Coverage)}}   & Baseline & \textbf{77.8} & \textbf{84.9} & 75.6 & 80.6 & 78.1 & \textbf{85.1} & 79.9 & 80.5\\
                                & EMBR & \textbf{77.8} & 84.7 & \textbf{75.7} & 81.6 & \textbf{78.4} & 85.0 & \textbf{80.3} & \textbf{80.9}\\
                                & SCD loss   & 77.3 & 84.8 & 75.6 & \textbf{81.9} & 78.3 & 84.0 & \textbf{80.3} & \textbf{80.8}\\
                                \bottomrule
\end{tabular}
}
\end{table*}

We extract 128-dim log-Mel filter-bank energies from a 32ms window, stack every 4 frames, and sub-sample every 3 frames, to produce a 512-dimensional acoustic feature vector with a stride of 30 ms as the input to the acoustic encoder. The audio encoder shares the same model architecture and hyper-parameters as in \cite{xia2022turn} except that in this work we change the ``Dense layer 2" to 512 dimensions to increase model capacity. For the label encoder, we use a single 128-dim LSTM layer. The text tokens are projected into a 64-dim embedding vector before feeding into the label encoder. For the joint network, we have a projection layer that projects both the audio and label encoder outputs to 512-dim. At the output of the joint network, it produces a distribution over 75 possible graphemes (the English alphabet, punctuation, the speaker change token \texttt{<st>}, and special symbols like ``\$") with a softmax layer. The model has around 27M parameters in total. For optimization, we follow the same hyper-parameters as described in~\cite{zhang2020transformer}. We evaluate three systems that share the same model architecture,

\begin{itemize}
    \item \textbf{Baseline}: Trained with the negative log probability loss.
    \item \textbf{EMBR}: Warm-started from the model parameters of the \textbf{Baseline} model and fine-tuned with a linear combination of the EMBR loss $L_{\text{EMBR}}$ (Eq. (8) in \cite{shannon2017optimizing} computed on the 4-best hypotheses) and the negative log probability. The final loss is formulated as $L_{\text{EMBR}}-0.03~\log P(\mat{Y}|\mat{X})$. This serves as another baseline.
    \item \textbf{SCD loss} (proposed): Warm-started from the parameters of the \textbf{Baseline} model and fine-tuned with the $L_{\text{SCD}}$ loss in Eqs. (\ref{eq:per_hyp_loss}) and (\ref{eq:batch_loss}), and computed on the 4-best hypotheses. Empirically, we set ${\alpha, \beta, \gamma, \lambda}$ to 1, 10, 10, 0.03, respectively. We set $k=1.1$ in Eq. (\ref{eq:ins_del_costs}).
\end{itemize}

\vspace{-3mm}
\section{Results and Discussion}
\label{sec:res}
\vspace{-1mm}
Due to space constraints\footnote{\scriptsize Supplemental results and resources can be found at \url{https://github.com/google/speaker-id/blob/master/publications/ScdLoss}}, here we only report the recall rates described in Sec. \ref{sec:precision_recall} based on interval counts. Note that recall rates based on the interval duration follow the same trend. We compute purity and coverage rates with the \textit{pyannote.metrics} package \cite{bredin2017pyannote}.

\vspace{-4pt}
\subsection{Long-form results}
\label{sec:res:longform}
\vspace{-4pt}
The evaluation results on the original long-form data are summarized in Table \ref{tab:long_results} and the metrics follow the definitions in Sec. \ref{sec:metrics}. The F1 score is the harmonic average of the two metrics involved. Overall, the three systems perform similarly on the purity-and-coverage F1 score, with the \textbf{EMBR} and \textbf{SCD loss} systems performing slightly better than the \textbf{Baseline}. On the precision-and-recall F1 score, the proposed \textbf{SCD loss} system outperforms the \textbf{Baseline} and \textbf{EMBR} systems by 8.9\% and 3.5\% relative, respectively. More specifically, the \textbf{SCD loss} improves the recall rate by 16.8\% relative compared with the \textbf{Baseline} system while maintaining a similar level of precision (only a -0.6\% relative regression). The DIHARD1, Inhound, and Outbound test sets do not have corresponding training sets, therefore, the precision-and-recall F1 score improvements on these test sets demonstrate that the proposed training technique can generalize to out-of-domain data.

\begin{table}[b!]
\centering
\caption{Short-form results by pooling all test sets together.}
\label{table:shortform}
\resizebox{3.25in}{!}{%
\begin{tabular}{cccc}
\toprule
Length                & System    & F1 (Precision \& Recall) & F1 (Purity \& Coverage) \\
                      \cmidrule(lr){1-1} \cmidrule(lr){2-2} \cmidrule(lr){3-4}
\multirow{3}{*}{30s}  & Baseline  & 55.2 & 75.9\\
                      & EMBR     & 60.9 & 80.8\\
                      & SCD loss & \textbf{65.0} & \textbf{81.5}\\
                      \cmidrule(lr){1-1} \cmidrule(lr){2-2} \cmidrule(lr){3-4}
\multirow{3}{*}{60s}  & Baseline  & 58.6 & 77.9\\
                      & EMBR     & 64.4 & \textbf{81.1}\\
                      & SCD loss & \textbf{67.9} & \textbf{81.1}\\
                      \cmidrule(lr){1-1} \cmidrule(lr){2-2} \cmidrule(lr){3-4}
\multirow{3}{*}{120s} & Baseline  & 61.8 & 79.5\\
                      & EMBR     & 66.6 & \textbf{81.2}\\
                      & SCD loss & \textbf{69.6} & 81.0\\
                      \bottomrule
\end{tabular}
}
\end{table}

For comparison, we also calculate the precision-and-recall F1 scores following their conventional definitions \cite{yin2017speaker} (not shown in Table \ref{tab:long_results}) using the implementation in the \textit{pyannote.metrics} package. We set the collar value to 250ms on each side (the same as what we used in our precision and recall calculation). Overall, the \textbf{SCD loss} system has an F1 score of 36.5, outperforming the \textbf{Baseline} (32.2) by 13.4\% relative, and the \textbf{EMBR} system (34.6) by 5.5\% relative. Although the relative quality difference across systems remain consistent with the metrics we define in this work, the absolute value of the \textit{pyannote.metrics} precision and recall metrics are not representative of the true system performance (as reflected by the coverage and purity measures) due to the small collar value, as discussed in Sec. \ref{sec:metrics}. Simply increasing the collar value (e.g, to $>$1s) would not solve this issue since that would make these metrics too lenient on errors.

\vspace{-6pt}
\subsection{Short-form results}
\label{sec:res:shortform}
\vspace{-2pt}
We are also interested in the models' performance on short utterances since a wide range of applications focus on recordings that are less than a couple of minutes in duration. Therefore, we segment the long-form data into shorter utterances with various target lengths (30s, 60s, 120s). Note that the segmentation was done based on speaker annotations to avoid chopping in the middle of a sentence. Results are summarized in Table \ref{table:shortform}. On the precision-and-recall F1 score, the \textbf{SCD loss} system performs the best across all conditions. On the purity-and-coverage F1 score, the \textbf{SCD loss} and \textbf{EMBR} systems perform similarly while outperforming the \textbf{Baseline} system. The \textbf{SCD loss} system performs better when the recordings are around 30s long, and the \textbf{EMBR} system performs slightly better when the segments are longer (120s). One possible explanation is that the training data is around 15s on average, so the \textbf{SCD loss} model is tuned towards performing better on shorter segments.

\vspace{-7pt}
\section{Conclusions}
\label{sec:conclusions}
\vspace{-3pt}
We introduce a novel token-based training loss that directly minimizes the SCD error rates for a Transformer-Transducer based SCD model. We also propose a new set of definitions for calculating the precision and recall rates for SCD evaluation. Experiments on a set of diverse evaluation sets demonstrate that the proposed training loss can significantly improve the recall rate of SCD while maintaining the precision rate. We show that the proposed new metrics can highlight model quality differences when the conventional purity and coverage scores cannot, hence providing additional insights for model improvements.

\vspace{-8pt}
\section{Acknowledgements}
\vspace{-2pt}
The authors would like to thank Jason Pelecanos, Hank Liao, Olivier Siohan, Fran\c{c}oise Beaufays, Pedro Moreno Mengibar, and Ha\c{s}im Sak for their help.

\clearpage
\bibliographystyle{IEEEbib}
\bibliography{refs}

\end{document}